\documentclass[aps,floats,floatfix,amssymb,amsmath,prb,twocolumn,nofootinbib,superscriptaddress]{revtex4-2}

\usepackage{calc}
\usepackage{psfrag}
\usepackage{graphicx}
\usepackage{soul,color}
\usepackage{enumerate}
\usepackage[dvipsnames]{xcolor}
\usepackage[unicode=true,pdfusetitle,
citecolor=blue,bookmarks=true,bookmarksnumbered=false,bookmarksopen=false,
breaklinks=false,pdfborder={0 0 0},backref=false,colorlinks=true]
{hyperref}
\usepackage{cleveref}
\usepackage{verbatim}
\usepackage{geometry}
\usepackage{float}

\setstcolor{blue}

\newcommand{\up}{\uparrow}
\newcommand{\dn}{\downarrow}



\begin{document}
	
	\title{Thermoelectric effect on diffusion in the two-dimensional Hubbard model}
	
	\author{Martin Ulaga}
	\affiliation{Jo\v{z}ef Stefan Institute, Jamova 39, 1000 Ljubljana, Slovenia}
	\author{Jernej Mravlje}
	\affiliation{Jo\v{z}ef Stefan Institute, Jamova 39, 1000 Ljubljana, Slovenia}
	\affiliation{University of Ljubljana, Faculty of Mathematics and Physics, Jadranska 19, 1000 Ljubljana, Slovenia}
	\author{Jure Kokalj}
	\affiliation{University of Ljubljana, Faculty of Civil and Geodetic
		Engineering, Jamova 2, 1000 Ljubljana, Slovenia} 
	\affiliation{Jo\v{z}ef Stefan Institute, Jamova 39, 1000 Ljubljana, Slovenia}
	
	\begin{abstract}
		We study charge and heat transport in the square lattice Hubbard model at strong coupling using the finite-temperature Lanczos method. We construct the diffusion matrix and estimate the effect of thermoelectric terms on diffusive and hydrodynamic time evolution. The thermoelectric terms prevent the interpretation of the diffusion in terms of a single time scale. We discuss our results in relation to cold-atom experiments and measurements of heat conductivity based on the measurements of heat diffusion.
	\end{abstract}
	\maketitle

	\section{Introduction}
	
	Strong correlations lead to unusual phenomena such as unconventional superconductivity~\cite{nguyen2021superconductivity,hayes2021superconductivity},  non-Fermi-liquid behavior~\cite{stewart2001,hill01}, strange metallicity~\cite{legros2019universal}, ND transport without  quasiparticles~\cite{pustogow2021rise,chen2022shot}, to name a few. Solutions of microscopic Hamiltonians provide crucial insights to aid the interpretation of experiments and guide phenomenological theory approaches~\cite{hartnoll2022rmp,chowdhury2022rmp}. Recently, numerical simulations of the Hubbard model successfully described the high-temperature ``bad-metal'' regime~\cite{kokalj17} and also reached the strange metal regime~\cite{huang19}.
	
	Parallel efforts of simulating model Hamiltonians in cold atoms have led to a remarkable advance~\cite{bloch2008many,altman2021quantum} as well. Recent highlights include the simulation of charge~\cite{brown19} and spin~\cite{nichols19} dynamics in the square lattice Hubbard model and the observation of thermalization and a crossover from diffusive to sub-diffusive dynamics at infinite temperature ($T$)~\cite{gardadosanchez19}. In these setups, the transport properties are usually determined indirectly~\cite{borup2015measuring} from observing the time evolution of a chosen initial state (e.g. a density wave) without reaching a steady state with a fixed current.
	
	A crucial aspect that can affect the interpretation of such time evolution is the fact that the dynamics are coupled, with diffusion involving several quantities, such as charge and heat, due to the finite thermoelectric effect away from particle-hole symmetry. Therefore, the discussion should account for the associated mixed dynamics~\cite{mravlje2022spin}. With cold atoms, the thermoelectric effect has been investigated for a gaseous system in the bottleneck geometry~\cite{brantut2013,krinner2017two,hausler2021interaction}; however, it has not been explored in optical lattices and was assumed to be negligible in the interpretation of existing lattice results.
	
\begin{figure}[t!]
		\centering
		\includegraphics[width=.99\linewidth]{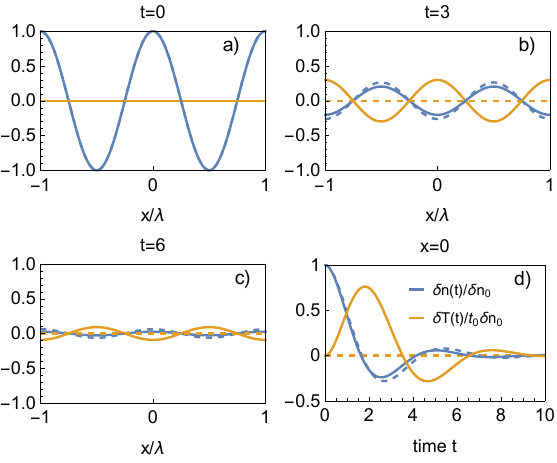}
		\caption{Snapshots of the time evolution of a charge density wave and accompanying temperature modulations are shown in (a) for the initial state, in (b) at time $t=3$, and in c) for $t=6$. (d) The time dependence of the wave amplitudes.  The modulation $\delta n(t)$ is normalized to the initial modulation amplitude $\delta n_0$, while the temperature modulation $\delta T(t)$ is normalized to $t_0 \delta n _0$, where $t_0$ is the hopping parameter. These results were obtained for the two-dimensional Hubbard model with $U=7.5t_0$ at 15\% hole doping and a temperature $T=t_0$. The initial state shown in (a) only contains the density modulation with a wavelength $\lambda=7a$ (in the $x$ direction) and has no temperature modulation, meaning that it is in thermal equilibrium. Dashed lines in the plots represent the results for the case where the thermoelectric effect is neglected. When accounting for the thermoelectric coupling, the density wave evolves differently over time, accompanied by the appearance of a temperature modulation.}
		\label{fig:snapshots-nxt}
	\end{figure}

	In this paper, we address the issue of mixed diffusion by considering the matrix diffusion equation. We calculate all needed quantities, including the ones related to the thermoelectric effect, in the square lattice Hubbard model using the finite-temperature Lanczos method (FTLM). We further use numerical results to obtain the hydrodynamic solution to the time evolution including current relaxation rates. As an example, in Fig.~\ref{fig:snapshots-nxt}, we show the solution of the coupled density-heat diffusion problem with diffusion matrix and current relaxation rates obtained from the numerical solution of the doped Hubbard model at a particular $T$. Due to the thermoelectric effect, the initial pure density profile additionally results in a $T$ profile as time evolves. The obtained time dependence differs from that when the thermoelectric mixing is neglected. In the Hubbard model at high $T$, accessible to our numerics, quantitatively the effect is moderate.  The density profile is seen to be close to the one obtained if the thermoelectric effects are neglected. We discuss why this is so and under what circumstances the effect can become larger. On the other hand, the emerging $T$ modulation is completely absent if thermoelectric effects are  neglected. 
	
	The qualitative aspects of our results apply not only to cold-atom experiments but also to measurements of diffusivity in general. One important example is a ``flash'' method, which  determines the heat conductivity from the propagation of the $T$ modulation~\cite{Parker1961}. More recent extensions of such a method, where the decay of a thermal wave introduced by periodic laser heating is studied, are also potentially affected by our considerations~\cite{zhang17,sun2023spatially}.
	
	Very recently, related calculations of the thermoelectric effect  were reported in Refs.~\cite{silva2023,wang23} that used the quantum Monte Carlo method on related lattice models. Whereas these remarkable state-of-the-art calculations reach large system sizes, the dynamical results rely on analytical continuation. It is important to cross-verify those results by a method that does not include the same systematic uncertainties (difficult to precisely quantify) and to estimate qualitatively and quantitatively  the effect of thermoelectric coupling on the time evolution for some typical experimental setups.
	
	This paper is structured as follows. We review the model and method, the hydrodynamic equations, and the diffusion matrix in Section~\ref{sec:modelmethod}. We present the impact of the thermoelectric effect on hydrodynamics in Section~\ref{sec:res} and discuss the implications for experiments in Section~\ref{sec:conc}. Appendix~\ref{sec:app_diffmatrix} contains details on the diffusion matrix, Appendix~\ref{sec:app_lij} contains details on the FTLM calculations, Appendix~\ref{sec:app_comp} gives a comparison of some quantities with results from other methods, Appendix~\ref{sec:app_swilson} contains more details on thermodynamic ratios, and Appendix~\ref{sec:app_gamma} contains further information on the extraction of lifetimes from correlation functions.
	
	\section{Methods}
	\label{sec:modelmethod}
	\subsection{The Hubbard model}
	
	We use the two-dimensional Hubbard model with the Hamiltonian
	\begin{equation}
		H=-\sum_{\langle ij\rangle \sigma}t_{ij}c^{\dagger}_{i\sigma}c_{j\sigma}+U\sum_{i}n_{i\up}n_{i\dn},
	\end{equation}\noindent
	where $t_{ij}$ is the hopping integral between nearest neighbors on a square lattice (we set $t_{ij}=t_0$) and $U$ is the local Hubbard interaction. We treat the model on a finite $4\times 4$ cluster using the FTLM~\cite{jaklic00} and avoid showing low-$T$ results affected by finite-size effects. See also Appendix~\ref{sec:app_lij} for more details on the method. We use $\hbar=k_B=e_0=g\mu_B=1$. When not written out explicitly, we use $t_0$ as the unit of energy and the lattice spacing $a$ as the unit of distance.
	
	\subsection{Transport coefficients}
	
	Gradients of $T$ and chemical potential $\mu$ induce currents  as given by the transport coefficients $L_{ij}$.
	\begin{align}
		j&=-L_{11}\nabla\mu-L_{12}\frac{\nabla T}{T},\\
		j_{q}&=-L_{12}\nabla\mu-L_{22}\frac{\nabla T}{T}.
	\end{align}
	The transport coefficients are related to charge and heat conductivities as
	\begin{align}
		&\sigma_c=L_{11}, &\kappa=\frac{1}{T}\left(L_{22}-\frac{L_{12}^2}{L_{11}}\right).
	\end{align}
	The Seebeck coefficient $S$ is the ratio between the gradient of voltage and the temperature gradient 
	\begin{equation}
		S=\frac{\nabla\mu}{\nabla T}=-\frac{L_{12}}{L_{11}T}.
	\end{equation}
	We compute $L_{ij}$ from current-current correlation functions as described in Appendix~\ref{sec:app_lij}.

	\subsection{Diffusion matrix}
	\label{sec:diffmatrix}
	
	Gradients of chemical potential $\mu$ and temperature $T$ induce gradients of density and entropy (assuming local equilibrium),
	\begin{align}
		\nabla n&=\chi_c \nabla \mu+\zeta \nabla T,\\
		T \nabla s&=T\zeta \nabla \mu+c_\mu \nabla T.
	\end{align}
	Here, $\chi_c$ is the charge susceptibility, $c_\mu$ is the specific heat at constant $\mu$ and $\zeta$ is the thermoelectric susceptibility, e.g.,  $\zeta=\partial_T n|_\mu$. See also Appendix~\ref{sec:app_diffmatrix}. Using these relations together with continuity  equations, we can write (Appendix~\ref{sec:app_diffmatrix}) the diffusion equation for $n$ and $T$ as   
	\begin{equation}
		\partial_{t}\begin{pmatrix}
			n \\ T
		\end{pmatrix}=\mathbf{D} \nabla^2\begin{pmatrix}
			n \\ T
		\end{pmatrix}.
	\end{equation}
	The diffusion matrix (in the basis of $n$ and $T$) reads
	\begin{align}
		\mathbf{D}&=\left(
		\begin{array}{cc}
			\frac{L_{11}}{\chi_c} & \frac{L_{12}}{T}-\frac{\zeta  L_{11}}{\chi_c} \\
			\frac{L_{12} {\chi_c}-\zeta  L_{11} T}{c_n\chi_c^2} & \frac{\zeta ^2 L_{11} T^2-2 \zeta  L_{12} T \chi_c +  L_{22} \chi_c^2}{c_nT \chi_c^2} \\
		\end{array}\right)\label{eq:dmatrix_Lij}\\ 
		&=\left(\begin{array}{cc}
			D_c & \pm\sqrt{\frac{c_n \chi_c D_\mathrm{corr} D_c}{T}} \\
			\pm\sqrt{\frac{D_\mathrm{corr} D_c T}{c_n \chi_c}} &  \widetilde{D_Q} 
		\end{array}
		\right).
		\label{eq:dmatrix}
	\end{align}
	On the diagonal one has $D_c$ and $\widetilde{D_Q}=D_Q+ D_\mathrm{corr}$, which are the charge and heat diffusion constants for cases with no temperature or density modulations, respectively. $D_c$ and $D_Q$ are the standard diffusion constants, related to the corresponding conductivities by the Nernst-Einstein equations $\sigma_c=D_c\chi_c$ and $\kappa=D_Qc_n$. Here, $c_n=c_{\mu}-\zeta^2T/\chi_c$ is the specific heat at fixed density. Note that the diagonal element $\widetilde{D_Q}$ differs from the standard heat diffusion constant $D_Q$ by $D_\mathrm{corr}$, which also expresses the off-diagonal elements. This parameter may be written as  
	\begin{equation}
		D_\mathrm{corr}=D_c\Tilde{W}(S^K-S)^2,
		\label{eq:dcorr}
	\end{equation}
	\noindent
	and is related to the difference of the Seebeck coefficient from its thermodynamic Kelvin approximation~\cite{peterson2010}  $S^K=\partial_T \mu|_n$, namely to $S^K-S$ [the sign of the off-diagonal elements in Eq.~\eqref{eq:dmatrix} equals the sign of $S^K-S$].  $D_\mathrm{corr}$ is also connected to a modified ``Wilson ratio''  $\Tilde{W}=T\chi_c/c_n$ with {\it charge} susceptibility $\chi_c$ in the place of the more standard spin susceptibility. 
	
	$D_\mathrm{corr}$ is the key quantity that controls the effect of thermoelectric mixing and in turn the deviations of the diffusion matrix eigenvalues
	\begin{equation}
		D_{\pm}=\frac{D_c +\widetilde{D_Q}}{2} \pm \sqrt{\left(\frac{D_c - \widetilde{D_Q}}{2}\right)^2 + D_c D_\mathrm{corr},}
		\label{eq:dpm_formula}
	\end{equation}
	from standard diffusion constants $D_c$ and $D_Q$. It is important to keep in mind that $D_\mathrm{corr}$ also changes the diagonal element $D_{TT}$ to $\widetilde{D_Q}=D_Q+ D_\mathrm{corr}$, as discussed above. The diffusion matrix was recently also discussed in related models for bad~\cite{mendezvalderrama2021} and strange~\cite{davison2017} metals.
	
	Finally, we note that the form of diffusion matrix depends on the chosen basis; for example, the occurrence of $D_c$ in the element ${D}_{nn}$ of $\mathbf{D}$ is characteristic of the $(n,T)$ basis. This simple expression is associated with the fact that if $\nabla T=0$, the particle current is given by $j=\sigma_c (-\nabla \mu) =\sigma_c/\chi_c (-\nabla {n})$, i.e., the standard Fick's law. Analogously, if one chooses chemical potential and heat $(\mu, Q)$ as the basis, one finds a simple form for the heat-heat element of the diffusion matrix $D_{QQ}= L_{22}/(c_\mu T)$. When not written otherwise, we refer to $\mathbf{D}$ and its elements in the basis of $(n,T)$. See Appendix~\ref{sec:app_diffmatrix} for more details.

	\subsection{Hydrodynamics of charge}
	
	Let us first discuss a typical measurement of diffusion in, e.g., cold atom experiments~\cite{brown19,nichols19}. One prepares an initial state with some density modulation via some spatially modulated external potential. Such a state is initially in equilibrium and has no temperature modulation or currents. Next, the external potential is switched off and the system is left to evolve freely, during which time the density modulation starts to decay. In the case of negligible thermoelectric coupling, the density modulation decays according to the diffusion equation $\partial_t n=D_c \nabla^2 n$ and current flows according to the Fick's law
	\begin{equation}
		j+D_c\nabla n=0.
	\end{equation}
	However, Fick's law  dictates that the  current appears instantly after the external potential is switched off and is instantly proportional to the density gradient, while in reality the current needs some time to develop. For this reason, one introduces the current relaxation rate $\Gamma_c$ and uses the improved hydrodynamic description~\cite{kadanoff63},
	\begin{equation}
		\partial_{t}j+\Gamma_c(j+D_c\nabla n)=0.
		\label{eq:curr_relax}
	\end{equation}
	This description has been previously discussed in the context of the Hubbard model at various values of $U$~\cite{vucicevic2022charge}. Together with the continuity equation and a spatial Fourier transform for a wave vector $k$, one obtains the second-order differential equation
	\begin{equation}
		\partial_{t}^2 n + \Gamma_c (\partial_{t} n + D_ck^2 n)=0.
		\label{eq:diff_eq_full}
	\end{equation}
	This is the ordinary damped harmonic oscillator equation, and its solution is 
	\begin{align}
		n(t)&=a\textrm{Re}\left[\cos(\Tilde{\omega}t+\phi)\right]e^{-\Gamma_c t/2},
		\label{eq:n(t)}
		\\
		\Tilde{\omega}&=\sqrt{\Gamma_c D_c k^2-\Gamma_c^2/4}.
	\end{align}
	Throughout this paper, we set the phase $\phi$ (for finite $\Gamma$ cases) in such a way that initially, no current is flowing, or $\partial_t n|_{t=0}=0$. Explicitly, we set $\phi=\arctan(-
	\Gamma_c/2 \tilde \omega)$. The prefactor $a$ determines the initial amplitude of modulation and we plot the modulations relative to this initial amplitude.
	
	The resulting $n(t)$ actually represents the modulation from equilibrium density $n$ and we therefore in the following denote it with $\delta n(t)$ for clarity. It is shown in Fig.~\ref{fig:snapshots-nxt} (dashed lines) with parameters corresponding to the Hubbard model at 15\% doping. Similar to the damped oscillator, the time dependence of the density modulation amplitude exhibits an underdamped regime with oscillations for $D_c k^2 > \Gamma_c/4$ (e.g., for larger values of $k$), and an overdamped regime without oscillations for small $k$. One recovers purely diffusive behavior with $e^{-D_c k^2 t}$ for $D_ck^2\ll \Gamma_c$, realized, e.g., in the $k\to 0$ limit.  
	
	\subsection{Matrix formulation of mixed diffusion}
	When thermoelectric effects are finite, density and heat diffusion are not independent and one has to extend the hydrodynamic treatment in a matrix formulation. We define the density and temperature modulation vector $\vec{v}=[\delta n(x,t)/\delta n_0,\delta T(x,t)/(t_0 \delta n_0)]$ with $\delta n(x,t)$ representing the density modulation difference from the uniform equilibrium density $n$, $\delta T(x,t)$ representing the temperature modulation from the equilibrium uniform $T$, and $\delta n_0$ representing the initial density modulation amplitude. With this we generalize Eq.~\eqref{eq:diff_eq_full} to matrix form:
	\begin{equation}
		\partial_{t}^2 \vec{v} + \mathbf{\Gamma} (\partial_{t} \vec{v} + \mathbf{D}k^2 \vec{v})=0.
		\label{eq:diff_eq_mixed}
	\end{equation}
	Here, $\mathbf{D}$ is the diffusion matrix and $\mathbf{\Gamma}$ is a matrix of relaxation rates. These are phenomenological parameters but can be related to the microscopic theory. To achieve this we introduce $\mathbf{D}(\omega)$ using $L_{ij}(\omega)$ in Eq.~\eqref{eq:dmatrix_Lij}, and then we diagonalize $\mathbf{D}$ for each $\omega$ and extract the corresponding eigenmodes relaxation rates $\Gamma_\pm$ as the width (half-width at half maximum) of $D_{\pm}(\omega)$. See also Appendix~\ref{sec:app_gamma}. The solution of Eq.~\eqref{eq:diff_eq_mixed} can then be expressed as
	\begin{equation}
		\Vec{v}(t)=a_+\Vec{v}_+f_+(t)+a_-\Vec{v}_-f_-(t).
		\label{eq:v_gamma_2x2}
	\end{equation}
	Here, $\Vec{v}_\pm$ are the corresponding eigenvectors  with diffusion constants $D_\pm$ and relaxation rates $\Gamma_\pm$. The form of $f(t)$ again corresponds to the solution of the damped harmonic oscillator and is that of Eq.~\eqref{eq:n(t)}, but with  $D_c$ and $\Gamma_c$ replaced with $D_\pm$ and $\Gamma_\pm$, respectively. Prefactors $a_+$ and $a_-$ depend on initial conditions. 
	
	\begin{figure*}[t]
		\centering
		\includegraphics[width=0.99\linewidth]{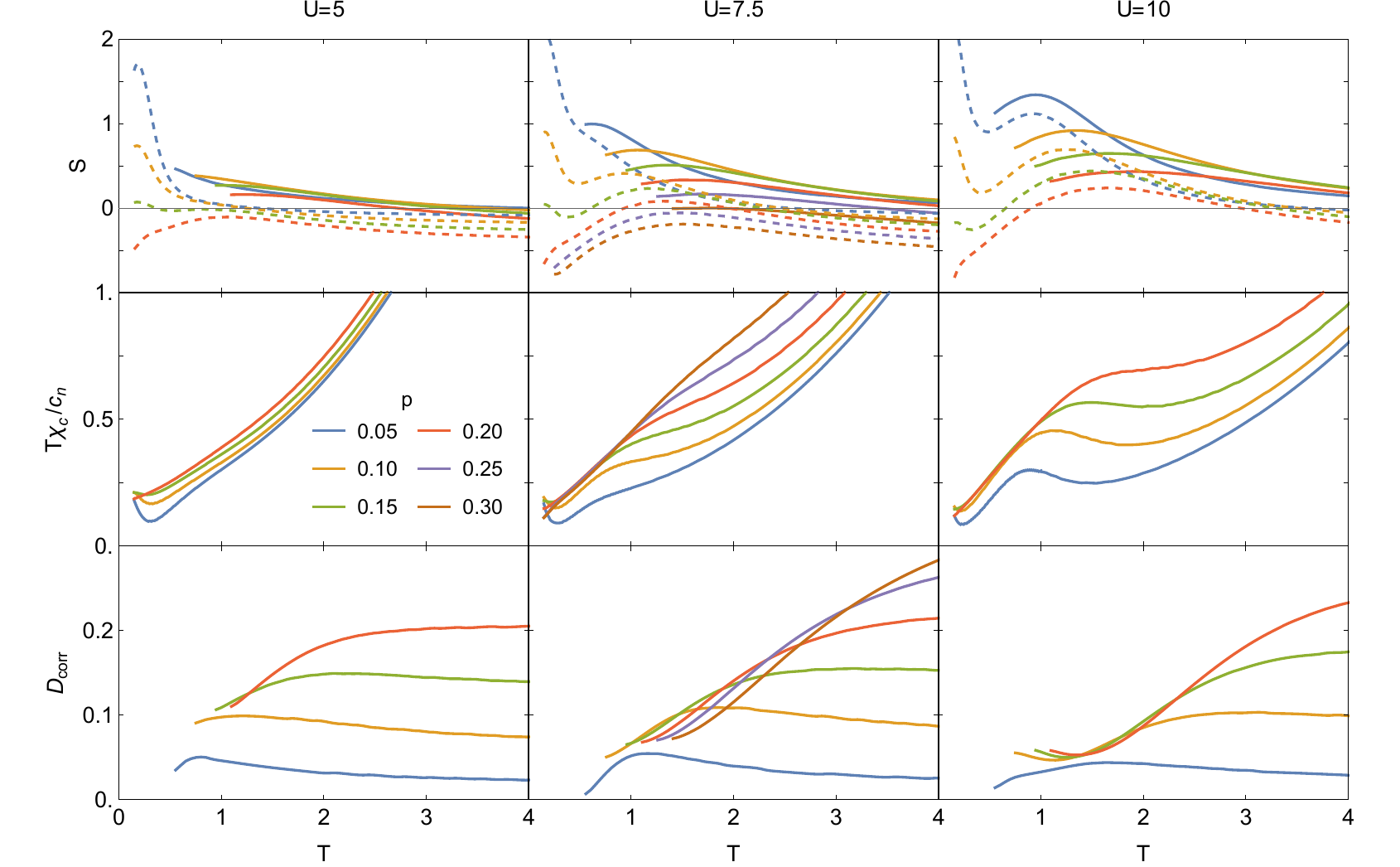}
		\caption{Top row: The Seebeck coefficient $S$ for the square lattice Hubbard model and for three interactions $U=5, 7.5$ and 10 and various dopings. The Kubo results (solid lines) are compared with the Kelvin approximation $S^K$ (dashed lines). Middle row: the ``Wilson ratio'' $\Tilde{W}=T\chi_c/c_n$. Bottom row: $D_\mathrm{corr}=D_c\Tilde{W}(S^K-S)^2$. }
		\label{fig:seebeck_allp}
	\end{figure*}

	\section{Results}
	\label{sec:res}
	\subsection{Hubbard model results}
	\label{sec:res_hubbard}
	Let us start with a discussion of the extent of the thermoelectric mixing, which is determined  by $D_\textrm{corr}$ and, via Eq.~(\ref{eq:dcorr}), by the deviation of the Seebeck coefficient $S$ from its Kelvin estimate $S_K$ and the modified Wilson ratio $\Tilde{W}$.
	
	In the top panels of Fig.~\ref{fig:seebeck_allp} we show the temperature dependence of $S$  (solid lines) and compare it with $S_K$ (dashed lines). In the considered regimes one expects $S$  to be characterized by a crossover from a high-temperature charge fluctuating regime characterized by the Heikes' \cite{chaikin76} value  $-\log[(2-n)/n] \approx -2 p$ (negative for hole doping $p$) to the regime with suppressed double occupancy (at large $U$ and small $T$) with Heikes' value $-\log[2(1-n)/n]$ (with positive values for considered $p$). One sees that these considerations indeed roughly describe the data.  With increasing $p$, the maximum in $S$ moves to higher $T$. $S$ increases moderately with increasing $U/t$ in a wide $T$ range. The Kelvin result suggests that $S$ changes sign  as a function of doping at $p\sim 0.15$ in the regime of lowest calculated $T$. Due to finite-size effects in the FTLM calculations at low $T$, we cannot observe this in the full Kubo calculation. Our results for $S$ and $S^K$ are qualitatively (for $S^K$ even quantitatively) consistent with the determinant quantum Monte Carlo (DQMC) results from Refs.~\onlinecite{wang23,silva2023} and we show a direct comparison also with our dynamical mean-field theory (DMFT) result in Appendix~\ref{sec:app_comp}.
	
	The key result for our discussion is that, despite considering a high-temperature regime ($T\sim 1$), we find that the difference $S^K-S$ is not small (one expects $S^K-S$ to drop as $1/T$ for $T \rightarrow \infty$) and approaches $k_B/e_0$ in the $U=10$ results.
	
	In the middle panels of Fig.~\ref{fig:seebeck_allp} we show the ``Wilson ratio'' $\tilde W$. The first observation is that the doping dependence is insignificant at $U=5$ but becomes more pronounced at larger $U$. At large $U$ and small doping, an additional intermediate peak develops. At high $T$, $\Tilde{W}=T\chi_c/c_n\sim T^2$ since $\chi_c\sim T^{-1}$ and $c_n \sim T^{-2}$. On lowering $T$, $\Tilde{W}$ drops and at larger interactions develops a plateau. At lowest $T$ and for small dopings, $\Tilde{W}$ grows again, which can be attributed to increased $\chi_c$~\cite{kokalj17, bonca03, brown19}. In the metallic Fermi-liquid regime at low $T$, one expects $\Tilde{W}$ to be $T$ independent.  Whereas in our simulations we cannot reach the Fermi-liquid regime due to the finite-size effects, we note that at our lowest $T$ the dimensionless quantity $\pi^2 \tilde W/3 \sim 0.5$ can be compared with the standard (spin) Wilson ratio (shown in Appendix~\ref{sec:app_swilson}) with values $\pi^2 T \chi_s/(3 c_n)\sim 2$. This points to a relatively increased spin susceptibility $\chi_s$ in comparison to $\chi_c$. The remaining weak dependence on $U$ with $\Tilde{W}$ that drops with $U$ at small $T$ can be rationalized as follows. To a first approximation $\chi_c = z g_0$ and $c_n = \pi^2 g_0 T/(3z)$~\cite{ulaga22}, where $g_0$ is the bare density of states at the chemical potential and $z$ is the quasiparticle weight. This leads to $\Tilde{W} = 3 z^2/\pi^2$, from which where one expects $\Tilde{W}$ to decrease with decreasing $z$ (increasing $U$). This decrease is indeed observed at the lowest calculated $T$.
	
	It is obvious from these results that neither $S^K-S$ nor $\Tilde{W}$ is particularly small and hence one does not expect $D_\mathrm{corr}$ to be negligible either. In the bottom panels of Fig.~\ref{fig:seebeck_allp} we show $D_\mathrm{corr}$. We see that this takes overall moderate values in our calculations (note that charge and heat diffusion constants are typically of order 1 at high $T$ ~\cite{ulaga22}). At highest $T$, $D_\mathrm{corr}$ tends to a constant because $(S^K-S)^2 \Tilde{W}$ and $D_c$ both become temperature independent there. At the lowest $T$ (not accessible in our calculations) in the Fermi-liquid regime one again expects a $T$-independent value of $D_\mathrm{corr}$ as $\Tilde{W} \to \mathrm{const}, S^K-S \propto T$, and  $D_c \propto 1/T^2$ there. We notice that $D_\mathrm{corr}/D_c\propto T^2 $ in the Fermi liquid and thermoelectric mixing has a limited effect at low $T$.

\begin{figure}[ht!]
		\centering
		\includegraphics[width=0.98\linewidth]{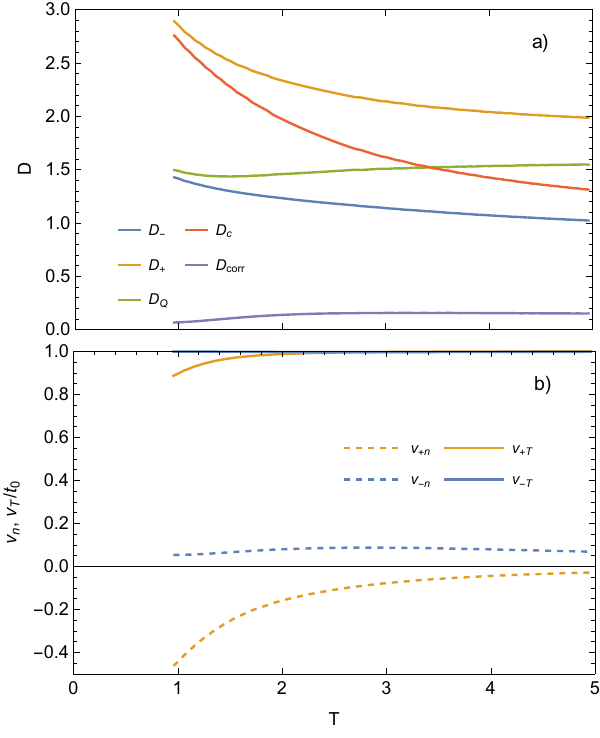}
		\caption{The temperature dependence of the eigenvalues of $\mathbf{D}$ compared with the bare diffusion constants (a) and their corresponding eigenvectors $\vec{v}_{\pm}$ (b). $D_\mathrm{corr}$ is also shown in the main text. Results are for the square lattice Hubbard model with $U=7.5$ and $p=0.15$.}
		\label{fig:dpm}
	\end{figure}

	We now consider a particular case of intermediate interaction $U=7.5$ and doping $p=0.15$. In Fig.~\ref{fig:dpm}(a), we show the bare diffusion constants $D_c, D_Q$ and the mixing element $D_\mathrm{corr}$, together with the diffusion eigenvalues $D_\pm$. One sees a growth of the charge diffusion constant on lowering $T$ and remarkably a much weaker temperature dependence of the heat diffusion constant $D_Q$ leading to a crossing of the two quantities at  $T\approx 3$, while no such crossing was observed for the case of spin and heat diffusion~\cite{mravlje2022spin}. The weaker temperature dependence and a shallow minimum of $D_Q$ are discussed in more detail in Ref.~\onlinecite{ulaga22}.
	
	The magnitude of $D_\mathrm{corr}$ is $\sim 10\%$ of the bare diffusion constants, leading to important effects of mixing when the two bare values are close. This is seen [Fig.~\ref{fig:dpm}(a)] from the temperature dependence of the two eigenvalues $D_\pm$ that follow a level-repulsion mechanism and hence differ significantly from the bare values.
	
	In Fig.~\ref{fig:dpm}(b) we show also the corresponding components of the eigenvectors. Looking at the $n$ components of the eigenvectors, one sees that at low $T$,  $\vec v_+$ has a larger $n$ component ($v_{+n}$). 
	At higher $T$, the larger $n$ component is in $\vec v_-$. 
	This is consistent also with the crossing of the bare diffusion constants. Furthermore, Fig.~\ref{fig:dpm}(b) shows that $n$ and $T$ components are in counter-phase for $\vec v_+$, while they are in phase for $\vec v_-$. Therefore, when the main component is $\vec v_+$ the $n$ and $T$ modulation are in counter-phase as, e.g., in Fig.~\ref{fig:snapshots-nxt}, while they are in phase when the $\vec v_-$ component is the dominant one. Which component dominates is determined by the initial condition via $a_{\pm}$ and the decay rate of each of the components.
	
	\subsection{Time evolution for mixed diffusion}
	\label{sec:res_eig}
	How important are the effects of mixing for the determination of diffusion constants from  the time evolution, such as is done in cold-atom experiments? We start the discussion assuming a fast relaxation limit $Dk^2\ll\Gamma/4$, e.g., due to the long-wavelength limit $k\to 0$. The time evolution in this limit is purely diffusive and is given by the matrix form of the diffusion equation and its solution 
	\begin{equation}
		\Vec{v}(t)= \exp(- \mathbf{D} k^2 t) \Vec{v} (0).
	\end{equation}
	It can be expressed also in terms of the eigenmodes 
	\begin{equation}
		\Vec{v}(t)=a_+ \Vec{v}_+ e^{-D_+k^2t}+ a_- \Vec{v}_- e^{-D_-k^2t},
		\label{eq:nqt}
	\end{equation}
	\noindent
	where $a_{\pm}$ are  coefficients set by the initial condition. Except in a special case where one of $a_{\pm}$ vanishes, the time evolution involves two time scales. 
	
	Let us consider the initial state $\Vec{v}(0)=(1,0)$ (pure density modulation) and ask about the density modulation at later times.  At short times, before appreciable temperature modulation develops,  $\delta n(t)$ falls as dictated by the diagonal entry $D_c$ of the diffusion matrix [Eq.~\eqref{eq:dmatrix}]. Alternatively, from the perspective of Fick's law, a pure density modulation drives the charge current given by $D_c$.  At long times, only the slower decaying eigenmode $\Vec{v}_-$ survives and the long-time dynamics are given by the corresponding eigenvalue $D_-$. 
	
\begin{figure}[t!]
		\centering
		\includegraphics[width=0.98\linewidth]{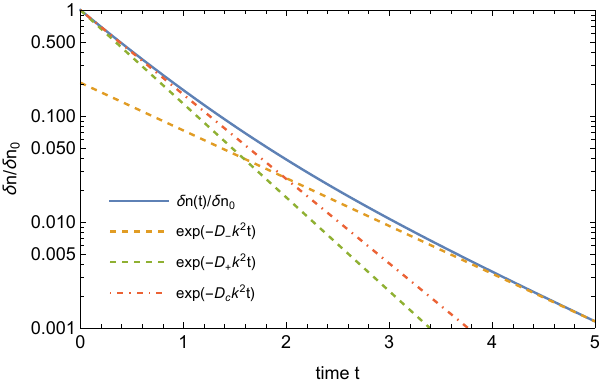}    
		\caption{The decay of a density modulation $\delta n$, including the thermoelectric effect (Eq.~\eqref{eq:nqt}). The full solution shows two time scales and interpolates in slope between {$\delta n\sim \exp(-D_ck^2t)$ for short times $t$ and $\delta n\sim \exp(-D_-k^2t)$ for long $t$}. The parameters are $U=7.5$, $T=1.5$, $k=2\pi/7$, and $p=0.15$.}
		\label{fig:timeevol_log}
	\end{figure}

	This behavior is illustrated in Fig.~\ref{fig:timeevol_log} which shows {$\delta n(t)$} for $U=7.5$ at $T=1.5$. There one sees that the solution begins to drop according to $\exp(-D_c k^2 t)$ (initial short time dependence $\propto 1-D_c k^2 t$ holds strictly) while at long times one sees exponential decay with time constant $(D_- k^2)^{-1}$. The full result is the sum of two exponentials. 
	
\begin{figure}[ht!]
		\centering  
		\includegraphics[width=0.99\linewidth]{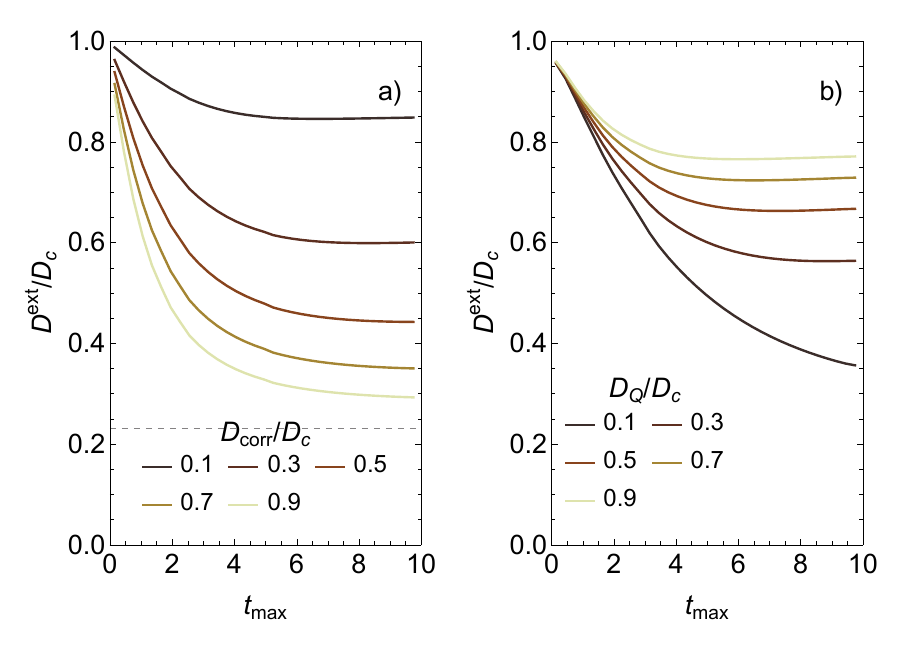}
		\caption{The effective diffusion constant as observed from the time evolution in an idealized setting by varying $D_\mathrm{corr}$ (a) and $D_Q$ (b). The effective diffusion constant $D^\textrm{ext}$ is obtained by fitting a single exponential up to $t_\mathrm{max}$. The fixed values of diffusion constants used are $D_Q/D_c=0.5$ for (a) and  $D_\mathrm{corr}/D_c=0.25$ for (b). The minimum possible $D^\mathrm{ext}=D_-$ is denoted with a dashed line in (a) for the case $D_\mathrm{corr}/D_c=0.9$. Here, $k=1$.} 
		\label{fig:timeevol_d2d4}
	\end{figure}

	In experiments, one often assumes a simple single exponential decay and fits the observed  time dependence with {$\delta n(t)=\delta n_0\exp(-D^\mathrm{ext} k^2 t )$}. It is now clear that the extracted diffusion constant $D^\mathrm{ext}$ depends on the fitting range. We illustrate this by showing $D^\mathrm{ext}/D_c$ for several values of $D_\mathrm{corr}$ as a function of the fitting range in Fig.~\ref{fig:timeevol_d2d4}, taking $D_Q/D_c =0.5$. One obtains sizable deviations of $D^\mathrm{ext}/D_c $ from 1 only for large values of $D_\mathrm{corr}$  and for longer fitting times. If the fitting range is very long, one approaches $D^\textrm{ext}\sim D_-$. One reaches $D^\textrm{ext} = D_-$ when  only the long-time regime is fitted and the short-time regime is left out.
	
	Since smaller $D_Q$ lowers $D_-$ to which $D^\textrm{ext}$ tends at longer fitting times, a smaller $D_Q$ also leads to a bigger mismatch and lower values of $D^\textrm{ext}/D_c$. Similarly, increasing $D_\textrm{corr}$ decreases $D_-$ via the level repulsion scenario and again leads to decreasing $D^\textrm{ext}/D_c$. These findings are summarized in Fig.~\ref{fig:timeevol_d2d4} and we note that the effect of $D_\mathrm{corr}$ is already significant at $D_\mathrm{corr}/D_c\sim 0.2$.
	
	All this illustrates that in principle the effects of thermoelectric coupling can be large and a na\"ive application of a bare diffusion with neglected thermoelectric effects can lead to a significant error in the estimate of the diffusion constant. On the other hand, it is reassuring, that at least at very short times, the decay rate is indeed governed by $D_c$. However, at such times, the current relaxation time can become important as discussed further in Sec.~\ref{sec:finitegamma}.

	\subsection{Finite-$\Gamma$ case and application to cold atom experiments}
	\label{sec:finitegamma}
\begin{figure}[ht!]
		\centering
		\includegraphics[width=0.98\linewidth]{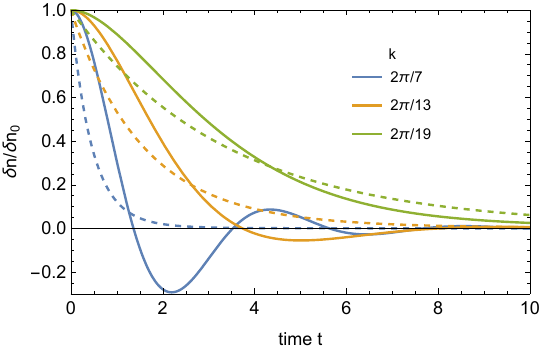}
		\caption{The time evolution of a density wave for an initial state with only density modulation. Solid lines are the full hydrodynamic solutions including the current relaxation rate [Eq.~\eqref{eq:v_gamma_2x2}]. Dashed lines are the solutions to the diffusion equation, ie. assuming $\Gamma\gg Dk^2$ [Eq.~\eqref{eq:nqt}]. The results are for several wave vectors $k$ and correspond to the {square lattice} Hubbard model with $U=7.5$ at $p=0.15$ and $T=1$.}
		\label{fig:timeevol_gamma}
	\end{figure}

	The measurements on optical lattices are performed with modulations with sizable momenta $k$ and hence one needs to take into account the current relaxation and keep $\mathbf{\Gamma}$ in Eq.~\eqref{eq:diff_eq_mixed} finite. The relaxation is estimated  as explained in Appendix~\ref{sec:app_gamma}. Snapshots of the resulting time evolutions are plotted in Fig.~\ref{fig:snapshots-nxt} (solid lines). In Fig.~\ref{fig:timeevol_gamma}, these are compared with diffusive solutions without current relaxation rates. The finite relaxation times lead to a slower decay at short times due to a slower initial buildup of currents, and to the oscillatory behavior as currents have some persistence and continue to flow even if the modulation becomes zero at a certain time.
	
	It is worth mentioning that each eigenmode decay is  determined by both $D_\pm$ and $\Gamma_\pm$ [Eq.~\eqref{eq:n(t)}]. Furthermore,  the eigenmode tends to exponential decay given with $e^{-D_\pm k^2 t}$ in the overdamped limit ($D_\pm k^2\ll \Gamma_\pm/2 $), while in the underdamped regime ($D_\pm k^2\gg \Gamma_\pm/2 $) it tends to oscillations suppressed with $e^{-\Gamma_\pm t/2}$. The long-lived mode is therefore given with the smaller value of $D_\pm$ in the overdamped (diffusive) regime, namely $D_-$ (as discussed above), while in the underdamped regime, it is given by the smaller value of {$\Gamma_{\pm}$}. It is possible that $\Gamma_+<\Gamma_-$ (as in the case as discussed in Appendix~\ref{sec:app_gamma}), making the longer lived mode in the underdamped regime $\vec v_+$ with corresponding out-of-phase modulation of $n$ and $T$ (see components in Fig.~\ref{fig:dpm}).  
	
	To estimate the impact of the thermoelectric effect in optical lattice measurements, we mimicked the analysis performed there. Namely, we obtain the solutions of the matrix hydrodynamic equations~\eqref{eq:diff_eq_mixed} which we fit with a simpler ansatz describing charge hydrodynamics~\eqref{eq:n(t)} only. We compared the results of this procedure to $D_c$ and $\Gamma_c$ obtained through FTLM calculations. 
	
\begin{figure}[t]
		\centering
		\includegraphics[width=0.98\linewidth]{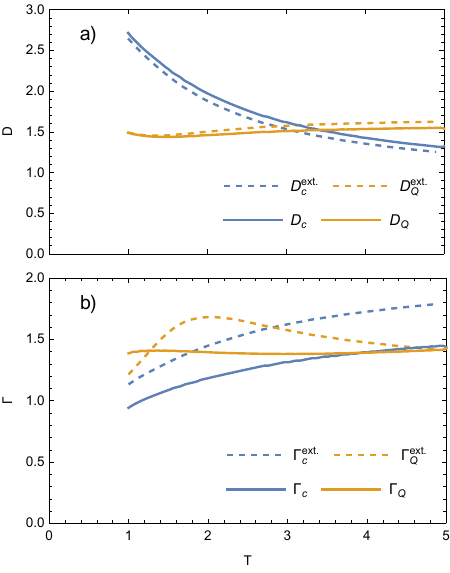}
		\caption{Comparison of the diffusion constant $D_c$ (a) and the scattering rate $\Gamma_c$ (b) as extracted by fitting Eq.~\eqref{eq:n(t)} to the first component of Eq.~\eqref{eq:v_gamma_2x2} up to times $t_\mathrm{max}=6t_0^{-1}$. Longer fitting time $t_\mathrm{max}=10t_0^{-1}$ affects $D_c^\mathrm{ext}$ and $\Gamma_c^\mathrm{ext}$ marginally.
			$D_Q$ and $\Gamma_Q$ are also shown. {Results are for the square lattice Hubbard model with $U=7.5$, $p=0.15$, and $k=2\pi/15$, although we checked that changing the wavevector does not alter the picture significantly.}}
		\label{fig:extracts}
	\end{figure}

	This analysis is summarized in Fig.~\ref{fig:extracts}(a). One sees that the extracted $D_c^\textrm{ext}$ is actually quite close to $D_c=\sigma_c/\chi_c$ in the entire temperature range. At low $T$ one could attribute this to a relatively large component  $|v_{+n}|$ and $D_+\sim D_c$. At $T$ where $D_c$ and $D_Q$ cross, $a_+v_{+n}\approx a_-v_{-n}$; that is, both eigenmodes are present in the initial state with similar weight and the mixing is close to maximal. Despite the fact that $D_+$ and $D_-$ are far from $D_c$, the initial time dependence is given by $D_c$ and extending the fitting time beyond $t_\textrm{max}=6 t_0^{-1}$ (with moderate $D_\textrm{corr}\sim 0.15$) results in $D^\textrm{ext}$ only slightly deviating from $D_c$. If one uses the value $D_c^\mathrm{ext}$ to calculate the resistivity via $\rho=\left(D_c^\mathrm{ext}\chi_c\right)^{-1}$, the estimation exceeds the value $\rho=\left(D_c\chi_c\right)^{-1}$ by $\sim10\%$. Figure~\ref{fig:extracts}(b) shows that, conversely, $\Gamma_c^\mathrm{ext}$ is not close to $\Gamma_c$ and is systematically overestimated. 
	
	\subsection{Effects of mixing on the thermal diffusion }
	The above considerations apply also to estimates of thermal transport based on measurements of thermal diffusion. The standard ``flash'' method estimates the thermal diffusion constant from the time it takes for the temperature on the back side of the sample to reach half of its equilibrium value after the front side has been illuminated by a laser pulse. It seems reasonable to assume that the initial state is described in terms of modulated temperature but that charge density is unaffected by the pulse; hence the diffusion matrix in the basis $(n,T)$ is appropriate to consider also in this case. Because the experimental procedure is sensitive to the initial time evolution before appreciable charge density gradients appear, the effects of the mixing with charge diffusion are expected to be limited.
	
	On the other hand,  it is important to recognize that the quantity obtained from such measurements is $\widetilde{D_Q}= D_Q + D_\mathrm{corr}$; hence if this quantity is used to estimate thermal conductivity using the Nernst-Einstein relation one obtains a diffusion estimate 
	\begin{equation}
		\kappa_\mathrm{diff}= \kappa (1 + D_\mathrm{corr}/D_Q)
	\end{equation}
	that is systematically larger from such measurements than what one obtains from the direct transport determination of $\kappa$.
	
	As a concrete example, we calculated $D_Q$ and $\Gamma_Q$ from a time evolution starting with a state containing a temperature modulation only. The results are shown in Fig.~\ref{fig:extracts} with dashed lines. Both $D_Q$ and $\Gamma_Q$ show deviations from $D_Q^\mathrm{ext}$ and $\Gamma_Q^\mathrm{ext}$. In particular, $\Gamma_Q$ is seen to be underestimated at low $T$ and overestimated the most at $T\sim 2t$ with the difference decreasing at higher $T$. $D_Q$ estimation is impacted differently from $D_c$ because of the occurrence of $D_\mathrm{corr}$ on the diagonal.
	
	We notice that if one assumes a different initial state, for instance with a constant chemical potential, the initial diffusion of temperature is given by the diagonal element in the $(\mu,T)$ basis,
	\begin{align}
		D_{TT}^{(\mu,T)}&=\frac{L_{22}\chi_c - \zeta L_{12} T}{c_nT\chi_c}\\
		&=D_Q+D_c \Tilde{W}S(S-S^K)
	\end{align}\noindent
	i.e., a value again distinct from standard diffusion $D_Q=\kappa/c_n$. Interestingly also here the deviations from $D_Q$ are given in terms of $S-S^K$ {and $\Tilde{W}$}.
	
	\section{Conclusions}
	\label{sec:conc}
	In conclusion, we investigated the mixed particle-heat diffusion in the doped Hubbard model. The thermoelectric effect caused the appearance of mixed diffusion modes of particles and heat and introduced new timescales that can alter the time dependence from that of the simple exponential decay. This should be taken into account in measurements in cold atom systems. We pointed out that the standard ``flash'' methods systematically give a higher value of thermal conductivity than what is obtained from the transport measurements (at least when the thermal conductivity is dominated by the electronic contribution). 
	
	It would be interesting to directly measure the mixed diffusion, for instance by introducing a temperature modulation into the system and studying the amplitude of the induced charge density wave. Because the dynamics are that of coupled damped oscillators, one for density modulation and one for $T$ modulation, one could also explore the resonating behavior as a function of driving frequency with the possibly enhanced dynamic thermoelectric effect.
	
	The effects of thermoelectric mixing are given by $D_\mathrm{corr}/D_c$.  
	This quantity was found to be moderate, $<\,0.2$, in our calculations but can become large in regimes where the charge susceptibility is large. For example, the divergence of $\chi_c$ in the vicinity of phase separation, such as in doped antiferromagnets~\cite{emery1993frustrated,kokalj17} or Hund's metals~\cite{demedici2017hund} enhances $\Tilde{W}$ and hence $D_\mathrm{corr}$. Such systems are good candidates to observe the predicted effects.
	
	\section*{Acknowledgment}
	We acknowledge support from the Slovenian Research Agency (ARIS) under Grant No. P1-0044 and J1-2458. JK and JM contributed equally to the work.
	
	\appendix

	\section{The diffusion matrix in the presence of spin fluctuations}
	\label{sec:app_diffmatrix}
	Here we give an overview of the diffusion matrix, which in general also includes spin properties, ie. is a $3\times 3$ matrix, even though we focus on a $2\times 2$ sub-block in the main text. The grand potential is given by
	
	\begin{equation}
		\Omega=E-{\cal{S}}T-\mu N-BM,
	\end{equation}
	\noindent
	where $\cal{S}$ is the entropy, $N=N_{\uparrow}+N_{\downarrow}$ is the number of particles and $M=(N_{\uparrow}-N_{\downarrow})/2$ is the magnetization. The entropy per site is given by
	\begin{equation}
		s={\frac{\cal{S}}{N_0}}=\frac{1}{N_0}\left(\log e^{-\beta\Omega}+\beta\langle \Hat{K}\rangle\right),
	\end{equation}
	\noindent
	where $\hat{K}=\hat{H}-\mu\hat{N}-B\hat{M}$ is the grand Hamiltonian, $\beta$ is the inverse temperature, and $N_0$ is the number of sites. Changes in density are described by
	\begin{align}
		dn&=-\frac{1}{N_0}\frac{\partial^2\Omega}{\partial\mu^2}d\mu-\frac{1}{N_0}\frac{\partial^2\Omega}{\partial\mu\partial T}dT-\frac{1}{N_0}\frac{\partial^2\Omega}{\partial\mu\partial B}dB\\
		&\equiv \chi_c d\mu+\zeta dT +\omega dB.\label{eq:dn}
	\end{align}
	Similarly, we have
	\begin{align}
		\begin{alignedat}{3}
			&\chi_s=-\frac{1}{N_0}\frac{\partial^2\Omega}{\partial^2 B},\quad &\xi=-\frac{1}{N_0}\frac{\partial^2\Omega}{\partial B\partial T},\quad & c_{\mu,B}=-\frac{T}{N_0}\frac{\partial^2\Omega}{\partial T^2},
		\end{alignedat}
	\end{align}
	\noindent
	and use these to express changes in entropy and magnetization. Together, these can be cast as a matrix equation~\cite{hartnoll15},
	\begin{align}
		\begin{pmatrix}dn\\Tds\\dm\end{pmatrix}=
		\begin{pmatrix}
			\chi_c & \zeta & \omega\\
			\zeta T & c_{\mu,B} & \xi T \\
			\omega & \xi & \chi_s
		\end{pmatrix}
		\begin{pmatrix}d\mu \\ dT\\ dB\end{pmatrix}
		\equiv\mathbf{A}\begin{pmatrix}d\mu \\ dT\\ dB\end{pmatrix}.
	\end{align}
	\noindent
	We use the Kubo formalism to obtain transport coefficients~\cite{shastry09} from transport equations {for particle, heat, and spin currents ($j$, $j_q$, and $j_s$, respectively).}
	\begin{align}
		\begin{pmatrix}
			j\\
			j_q\\
			j_s
		\end{pmatrix}=\begin{pmatrix}
			-L_{11} & -\frac{L_{12}}{T} & -L_{13}\\
			-L_{21} & -\frac{L_{22}}{T} & -L_{23}\\
			-L_{31} & -\frac{L_{32}}{T} & -L_{33}\\
		\end{pmatrix}\begin{pmatrix}
			\nabla\mu\\
			\nabla T\\
			\nabla B
		\end{pmatrix}\equiv \mathbf{L}\begin{pmatrix}
			\nabla\mu\\
			\nabla T\\
			\nabla B
		\end{pmatrix}
		\label{eq:lmatrix}.
	\end{align}
	\noindent
	Here, $L_{ij}=L_{ji}$, by Onsager reciprocity. 
	{Onsager reciprocity relations are valid even for finite frequencies~\cite{shastry09} and rely on time reversibility. Despite our hydrodynamic description in Eq.~\eqref{eq:diff_eq_mixed} involving the current relaxation rate, which breaks time-reversal symmetry, the underlying microscopic dynamics and Hamiltonian are time-reversal invariant, rendering Onsager's relations valid. 
	}
	{$j_q$ is related to the energy current $j_{\varepsilon}$ as}
	\begin{equation}
		j_q=j_{\varepsilon}-\mu j-Bj_s.
		\label{eq:heatcurr}
	\end{equation}
	Combining the above equations with continuity equations for conserved quantities
	\begin{align}
		\partial_{t} n + \nabla\cdot j=0,\nonumber\\
		\partial_{t} \varepsilon + \nabla\cdot j_\varepsilon=0,
		\label{eq:continuity}\\
		\partial_{t} m + \nabla\cdot j_s=0,
		\nonumber
	\end{align}
	\noindent
	one obtains a matrix-form diffusion equation
	\begin{align}
		\begin{pmatrix}
			\partial_t n\\T\partial_t s\\\partial_t m
		\end{pmatrix}=\mathbf{D}_0
		\begin{pmatrix}
			\nabla^2{n}\\T\nabla^2 s\\\nabla^2{m}
		\end{pmatrix},
		\label{eq:app_d0}
	\end{align}
	\noindent
	where $\mathbf{D}_0=-\mathbf{LA}^{-1}$ is the diffusion matrix {and also $dQ=Td{\cal S}$ is used}. 
	Using energy density $d\varepsilon$ is sometimes preferred to entropy density. In this case, the susceptibility matrix that enters is $\mathbf{\Tilde{A}}=\mathbf{P}_{\mu\varepsilon}\mathbf{A}$, where
	\begin{align}
		\mathbf{P}_{\mu\varepsilon}=\begin{pmatrix}
			1 & 0 & 0\\
			\mu & 1 & B\\
			0 & 0 & 1
		\end{pmatrix}.
	\end{align}
	To get the energy current, one multiplies Eq.~\eqref{eq:app_d0} from the left with $\mathbf{P}_{\mu\varepsilon}$ once more, arriving at
	\begin{align}
		\begin{pmatrix}
			\partial_t n\\\partial_t\varepsilon\\\partial_t m
		\end{pmatrix}=-\mathbf{P}_{\mu\varepsilon}\mathbf{LA}^{-1}\mathbf{P}_{\mu\varepsilon}^{-1}
		\begin{pmatrix}
			\nabla^2{n}\\\nabla^2 \varepsilon\\\nabla^2{m}
		\end{pmatrix}.
		\label{eq:app_deps}
	\end{align}
	To obtain the form of the diffusion matrix in the basis $(n,T)$ given in the main text, one uses 
	\begin{align}
		\mathbf{P}_{nT}^{-1}=\begin{pmatrix}
			1 & 0 \\
			\frac{T\zeta}{\chi_c} & c_n
		\end{pmatrix}
	\end{align}
	\noindent
	with the upper left $2\times 2$ block of $\mathbf{D}_0$. Then, $\mathbf{P}_{nT}\mathbf{D}_0\mathbf{P}_{nT}^{-1}$ gives the diffusion matrix given in the main text under Eq.~\eqref{eq:dmatrix}. Here the specific heat at constant density $c_n=c_{\mu}-T\zeta^2/\chi_c$ is used. The transformation into the basis of $(\mu,Q)$ taking $dQ=Td\cal{S}$ is achieved as $\mathbf{P}_{\mu Q}\mathbf{D_0}\mathbf{P}_{\mu Q}^{-1}$, where 
	\begin{equation}
		\mathbf{P}_{\mu Q}^{-1}=\begin{pmatrix}
			\chi_c-\frac{T\zeta^2}{c_{\mu}} & \frac{\zeta}{c_{\mu}}\\
			0 & 1
		\end{pmatrix}
	\end{equation}
	Finally, the transformation into the $(\mu,T)$ basis is the combination of the previous two,
	\begin{equation}
		\mathbf{P}_{\mu T}^{-1}=\begin{pmatrix}
			\chi_c & \zeta\\
			T\zeta & c_{\mu}
		\end{pmatrix}.
	\end{equation}
	Notice that the physics is contained in the eigenvalues of $\mathbf{D}$, which do not depend on the ``basis'' of $\mathbf{D}$. We use $T$ and $n$ as they are commonly used and experimentally monitored quantities.

	\section{Details of the FTLM calculation}
	\label{sec:app_lij}
	
	The transport coefficients $L_{ij}$ within the Kubo formalism are given by the $\omega\to 0$ limit of current-current correlation functions $L_{ij}(\omega)$, namely
	\begin{equation}
		L_{ij}(\omega)=\frac{1}{\omega N_0 V_\mathrm{u.c.}}\mathrm{Re}\int_0^{\infty}dte^{i\omega t}\langle [\hat{J}_i(t),\hat{J}_j(0)]\rangle.
		\label{eq:transport_coeff}
	\end{equation}
	We consider the particle, spin, and heat currents only in the $x$ direction. We have
	\begin{align}
		\hat{J}_n=&-it\sum_{j,\sigma,\delta} R_\delta^x c^{\dagger}_{j+\delta,\sigma}c_{j,\sigma},\\
		\hat{J}_s=&-it\sum_{j,\sigma,\delta} R_\delta^x \sigma c^{\dagger}_{j+\delta,\sigma}c_{j,\sigma},\\
		\hat{J}_\textrm{E}=&-\frac{it^2}{2}\sum_{j,\sigma,\delta,\delta '} R_{\delta\delta'}^x c^{\dagger}_{j+\delta+\delta',\sigma}c_{j,\sigma}\nonumber 
		\\&+\frac{itU}{2}\sum_{j,\sigma,\delta}R_\delta^x c^{\dagger}_{j+\delta,\sigma}c_{j,\sigma}(n_{j+\delta,\bar \sigma}+n_{j,\bar \sigma}),\\
		\hat{J}_\textrm{Q}=&\hat{J}_\textrm{E}-\mu\hat{J}_n-B\hat{J}_s,
	\end{align}
	\noindent
	where $R_{\delta}^x=x_{j+\delta}-x_j$ and $R_{\delta\delta'}^x=x_{j+\delta+\delta'}-x_j$ ($\delta$ point to the nearest neighbors of site $j$). We evaluate Eq.~\eqref{eq:transport_coeff} and thermodynamic quantities on a $4\times 4$ cluster using FTLM. {Within FTLM one averages over initial random vectors that are expressed with approximate Lanczos eigenvectors. These are then used to calculate static and dynamic quantities. The dynamic quantities require another set of Lanczos eigenvectors to calculate the current's matrix elements and spectral representation of the dynamic quantity. For more details see Refs. ~\cite{jaklic00, prelovsek13,  kokalj13}. We also use  averaging over twisted boundary conditions \cite{poilblanc91, bonca03} or averaging over shifts of wave vectors in the Brillouin zone, which further reduces the finite-size effect and is, e.g., able to reproduce the thermodynamic result for $U=0$ correctly.} {Finite-size effects in dynamic quantities also appear as a finite stiffness or a finite delta function at zero frequency, e.g., in optical conductivity (see, for example, Eq.~5.3 in Ref.~\onlinecite{jaklic00}). This appears to be due to particles crossing cluster boundaries without scattering, while stiffness should be zero at finite $T$ for normal (non superconducting, non integrable) dissipating systems in the thermodynamic limit. In our calculations, finite and large stiffness appears at low $T$ as a finite-size effect. We do not show low-$T$ regimes where the weight of this zero-frequency delta function exceeds $0.1\%$ of the total spectral weight $\int L_{ij}(\omega)d\omega$. The size of the stiffness is related to the spectral sum rule and diagonal matrix elements \cite{jaklic00,shastry06} and to the variation of energies with phase \cite{castella95}.
	}
	
	\section{Comparison with other computational techniques}
	\label{sec:app_comp}
	
\begin{figure*}[ht]
		\centering
		\includegraphics[width=0.49\linewidth]{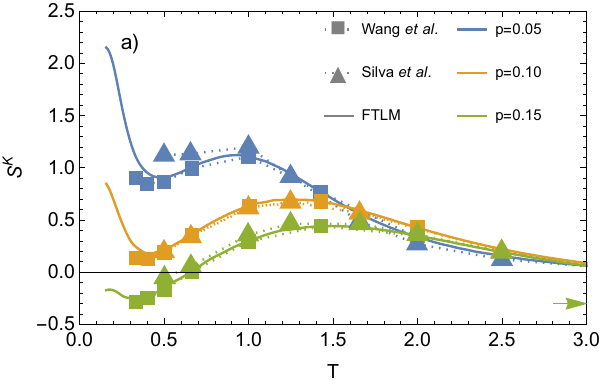}
		\includegraphics[width=0.49\linewidth]{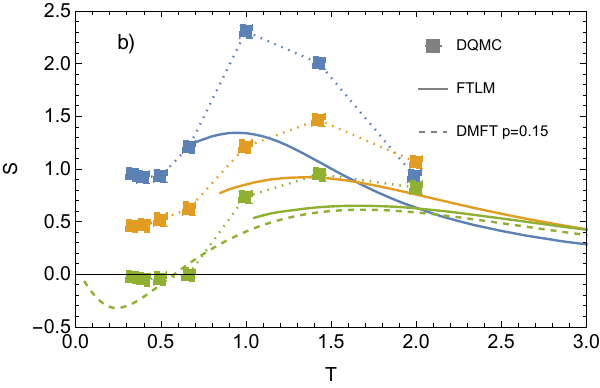}
		\caption{{(a) The temperature dependence of the Kelvin approximation for the Seebeck coefficient $S^K$ as calculated with our FTLM and  compared with the DQMC results of Silva \textit{et al.}~\cite{silva2023} (triangles) and Wang \textit{et al.}~\cite{wang23} (squares). The precise values of $p$ for the data from Silva \textit{et al.}~\cite{silva2023} are 0.04, 0.10, and 0.14. The Heikes high-temperature value $-\log[(1+p)/(1-p)]$ is also shown with a green arrow for $p=0.15$. (b) The Seebeck coefficients obtained with the Kubo formula and our FTLM compared with the DQMC results from Wang \textit{et al.}~\cite{wang23} and our DMFT results for $p=0.15$.}
		}
		\label{fig:seeb_comp}
	\end{figure*}

	{It is instructive to compare our results for thermopower with other techniques, namely the available DQMC data~\cite{silva2023,wang23} and, additionally, single-site DMFT. We compare the Kubo and Kelvin results for the Seebeck coefficient in Fig.~\ref{fig:seeb_comp}. The static results show good agreement in the entire $T$ regime. The Kubo formula result of the DQMC calculation of Ref.~\cite{wang23} gives somewhat bigger values for $S$ than our FTLM calculation, particularly at intermediate $T\sim 1.5$, while at lower $T$ the disagreement seems smaller. Our DMFT data for $p=0.15$ are also surprisingly close to the FTLM result despite the marked difference in known results for both resistivity~\cite{vranic20} and thermal conductivity~\cite{ulaga22}. The difference is attributed to vertex corrections~\cite{vucicevic19} and their effect seems to somewhat cancel in $S$. Similar cancellation of vertex corrections was previously observed in the Lorenz ratio~\cite{ulaga22}. 
	}
	
	\section{The Wilson ratio}
	\label{sec:app_swilson}
	
\begin{figure*}[ht!]
		\centering
		\includegraphics[width=0.99\linewidth]{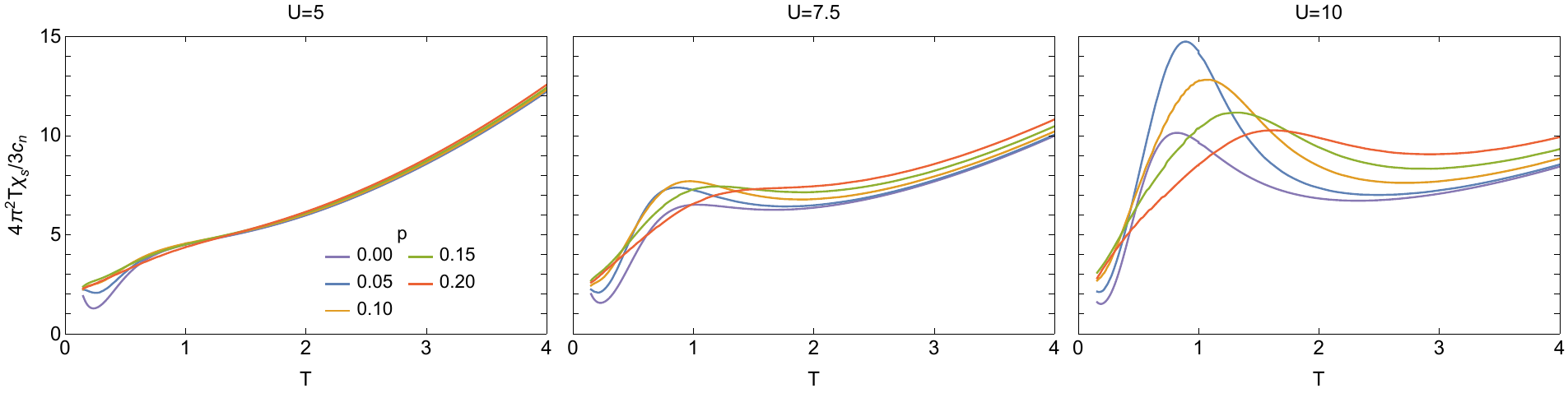}
		\caption{
			{The Wilson ratio $W$ as a function of temperature for various $U$ and dopings $p$.}}
		\label{fig:swilson}
	\end{figure*}

	{ It is interesting to compare the behavior of the ``Wilson ratio'' $\Tilde{W}=T\chi_c/c_n$ with that of the usual Wilson ratio involving spin susceptibility $\chi_s$,
		\begin{equation}
			W=\frac{4\pi^2 T\chi_s}{3c_n}. 
		\end{equation}
		Note that our definitions for $\Tilde{W}$ and $W$ differ by the factor $4\pi^2/3$. We show $W$ in Fig.~\ref{fig:swilson} as a function of $T$ for various $p$ and $U$. $W$ becomes  $\sim\,2$ at the lowest calculated $T$ and is only moderately dependent on $U$, which seems to be observed more generally \cite{vollhardt84}. As far as doping is concerned, one should distinguish the weak-coupling regime, where the dependence on doping is expected to be weak, and the strongly coupled doped Mott-insulator regime, where the correlations and magnetic susceptibility are expected to depend strongly on doping. Consistent with these expectations, we find that for $U=5$  the data at all dopings $p$ are essentially on top of each other. Increasing $U$ mainly has the effect that the $p$ dependence becomes more apparent, particularly in the peak located at $T\sim 1$. At high $T$, both $\chi_s$ and $\chi_c$ are proportional to $1/T$, and therefore both $W$ and $\Tilde{W}$ are proportional to $T^2$. Both quantities also develop a ``plateau'' at intermediate $T$ for larger $U$. 
		The differences between $W$ and $\Tilde{W}$ can be understood by comparing $\chi_c$ and $\chi_s$~\cite{kokalj17}. $W$ has been previously investigated in the $t-J$ model~\cite{jaklic00} (we note that the ratio reported in Ref.~\onlinecite{jaklic00} involves entropy instead of specific heat). 
	}
	
	\section{Details on extracting $\Gamma_{\pm}$}
	\label{sec:app_gamma}
	
	\begin{figure*}[ht]
		\centering
		\includegraphics[width=0.49\linewidth]{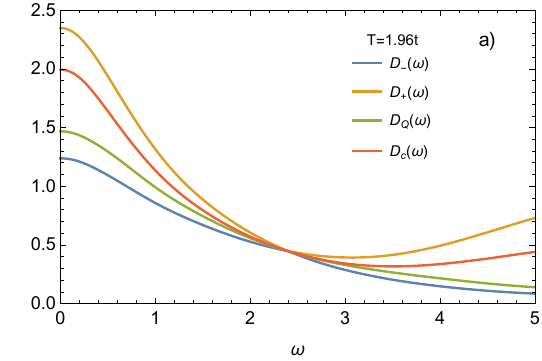}
		\includegraphics[width=0.49\linewidth]{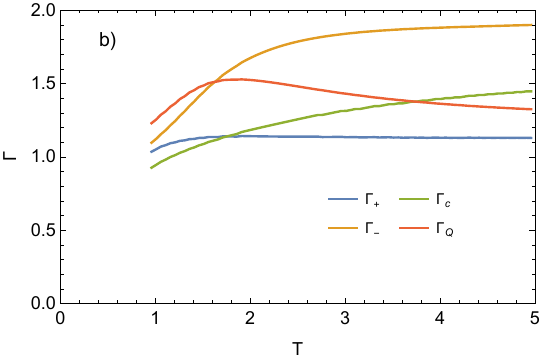}
		\caption{(a) The frequency dependence of diffusion constants generalized to finite $\omega$ $D_c(\omega)=
			\sigma_c(\omega)/\chi_c$, $D_Q(\omega)=\kappa(\omega)/c_n$, and $\bf D(\omega)$  eigenvalues $D_\pm(\omega)$. (b) The temperature dependence of the eigenmode scattering rates, also compared with the values extracted from $\sigma_{c}(\omega)$ and $\kappa(\omega)$, denoted with $\Gamma_{c}$ and $\Gamma_{Q}$.  {Results are for the square lattice Hubbard model with $U=7.5$, $p=0.15$, and $k=2\pi/15$ as in the main text.}}
		\label{fig:taupm_dpm_w}
	\end{figure*}
	
	Assuming a Drude form for the low-frequency part of dynamical conductivity, one can extract a scattering rate $\Gamma^0_c$ as the half-width of $\sigma_c(\omega)=L_{11}(\omega)$\footnote{Similarly, one can define $\Gamma_{\varepsilon}$ and $\Gamma_{c\varepsilon}$ from $\sigma_{\varepsilon\varepsilon}(\omega)$ and $\sigma_{c\varepsilon}(\omega)$.}.
	In the matrix generalization of $\mathbf{\Gamma}$, one has to account for additional relaxation rates. Just as $\Gamma_c$ can be obtained from the width of Drude peak of $L_{11}(\omega)$, one could determine elements of $\mathbf{\Gamma}$ from the $\omega$-widths of low-$\omega$ parts of $L_{ij}(\omega)$. We, however, use a slightly different approach and assume that the Drude-like form
	\begin{equation}
		\sigma_c(\omega) = \frac{\sigma_c(0)}{1-i\omega/\Gamma_c},\quad \kappa(\omega)=\frac{\kappa(0)}{1-i\omega/\Gamma_Q}
	\end{equation}
	\noindent
	is also applicable to $D_{\pm}(\omega)$ (i.e., generalized to finite $\omega$) for small $\omega$.  We therefore first calculate $\mathbf{D}(\omega)$ and obtain $\Gamma_{\pm}$ as the width of its eigenvalues $D_{\pm}(\omega)$,
	\begin{equation}
		D_{\pm}(\omega)=\frac{D_{\pm}(0)}{1-i\omega/ \Gamma_{\pm}}.
		\label{eq:d_omega}
	\end{equation}
	We find that the eigenvectors $\Vec{v}_{\pm}$ show weak enough frequency dependence in the considered regime at small $\omega$ that the obtained $\Gamma_{\pm}$ correspond to the $\omega=0$ diffusion matrix eigenvectors. At half filling, $\Gamma_{\pm}$ coincide with $\Gamma_c$ and $\Gamma_Q$ due to the vanishing thermoelectric effect.
	
	We show $D_{\pm}(\omega)$ in Fig.~\ref{fig:taupm_dpm_w}(a) where one sees that they indeed inherit the shape of the conductivities, justifying Eq.~\eqref{eq:d_omega} for small $\omega$, and that the general picture of ``level repulsion'' applies in the whole frequency range. Note that in the case of a $2\times 2$ diffusion matrix, $D_{\pm}(\omega)$ are guaranteed to be smooth functions, as evident from their closed-form expressions [Eq.~\eqref{eq:dpm_formula}].
	
	The frequency dependence of $D_{\pm}(\omega)$ reveals a feature at $\omega\approx 2.5 t$, where the various diffusion constants touch because $D_\textrm{corr}(\omega)=0$ and $S^K-S(\omega)$ changes sign. This occurs at $\omega$ exceeding $\Gamma_{\pm}$ and thus does not impact our estimates for $\Gamma_{\pm}$. Figure ~\ref{fig:taupm_dpm_w}(b) shows how $\Gamma_{\pm}$ differ from the bare values as obtained from $\sigma_c(\omega)$ and $\kappa(\omega)$, and one sees that they generically reinforce the ``level-repulsion'' picture at least at high $T$, where $\Gamma_+$ is about half of $\Gamma_-$. $\Gamma_{\pm}$ are decreasing in magnitude at low $T$, similarly to how $\Gamma_{c,Q}$ are expected to decrease as one approaches the coherent regime.

	
	%

\end{document}